\documentclass[10pt,twocolumn,showpacs,preprintnumbers,amsmath,amssymb,aps,prl,superscriptaddress,longbibliography]{revtex4-2}
\usepackage{appendix}
\usepackage{bbm}
\usepackage{mathrsfs}
\usepackage{graphicx}
\usepackage{dcolumn}
\usepackage{bm}
\usepackage{braket}
\usepackage{amsmath}
\usepackage{amsfonts}
\usepackage[dvipsnames]{xcolor}
\usepackage[colorlinks=true,linkcolor=Blue,urlcolor=BlueViolet,citecolor=BlueViolet]{hyperref}
\usepackage{natbib}
\usepackage{multirow}

\begin{document}

\title{Edge optical effect as a probe of chiral topological superconductors}
\author{Linghao Huang}
\affiliation{State Key Laboratory of Surface Physics and Department of Physics, Fudan University, Shanghai 200433, China} 
\affiliation{Shanghai Research Center for Quantum Sciences, Shanghai 201315, China}
\author{Jing Wang}
\thanks{wjingphys@fudan.edu.cn}
\affiliation{State Key Laboratory of Surface Physics and Department of Physics, Fudan University, Shanghai 200433, China}
\affiliation{Shanghai Research Center for Quantum Sciences, Shanghai 201315, China}
\affiliation{Institute for Nanoelectronic Devices and Quantum Computing, Fudan University, Shanghai 200433, China}
\affiliation{Hefei National Laboratory, Hefei 230088, China}

\begin{abstract}
We study the optical effect of chiral topological superconductors in two dimensions. The linear optical response from chiral Bogoliubov edge modes in clean superconductors has in-gap resonances, which is originated from the transitions within the edge particle-hole pair bands. Interestingly, the number of resonance peaks is determined by the Bogoliubov-de Gennes Chern number of topological superconductors. Such a sharply distinctive feature in optical absorption offers a simple way to distinguish topological superconductors with different Chern numbers. We further demonstrate that linear optical effect could probe different topological phases in quantum anomalous Hall insulator-superconductor junction devices. This finding provides an applicable method to detect chiral Bogoliubov edge states and is distinguishable from collective modes in superconductors and possible trivial explanations.
\end{abstract}

%\date{\today}

\maketitle

The chiral Bogoliubov edge modes are hybridized electron and hole states which propagate unidirectionally along the edge of two-dimensional topological matter~\cite{moore1991nonabelions,read2000paired,kitaev2006,fu2008superconducting,sau2010generic,alicea2010majorana,qi2011topological,wilczek2009majorana,elliott2015colloquium}. They have been attracting growing interests of study, for their potential applications in quantum computation~\cite{kitaev2003faulttolerant,nayak2008nonabelian,mong2014universal,clarke2013exotic,clarke2014exotic,lian2018topological,hu2018fibonacci}. A paradigmatic example hosting chiral Bogoliubov edge modes is the chiral topological superconductor (TSC), which can be realized in quantum Hall (QH) or quantum anomalous Hall (QAH)~\cite{chang2013experimental,Checkelsky_2014,kou2014, chang2015highprecision,mogi2015,deng2020quantum} states in proximity with an $s$-wave superconductor (SC)~\cite{qi2010chiral,wang2015chiral,chung2011conductance,wang2016electrically,strubi2011interferometric}. Prominent theoretical~\cite{fu2009probing,akhmerov2009electrically,qi2010chiral,wang2015chiral,chung2011conductance,wang2016electrically,strubi2011interferometric,lian2016edgestateinduced,lian2018,wang2018multiple,lian2019distribution,hu2024resistance,he2019platform,Hoppe2000,Giazotto2005,Akhmerov2007,Ostaay2011} and experimental~\cite{Wan2015,Amet2016,Lee2017,Sahu2018,Matsuo2018,Seredinski2019,Zhao2020,Zhao2023,Yuval2022,Hatefipour2022,uday2024,wang2024,vignaud_2023} efforts have been made towards realizing coherent chiral Bogoliubov edge modes, which allow coherent Andreev interference with ubiquitous transport signatures and offer possibilities for their coherent manipulation. However, the transport experimental signatures of chiral Bogoliubov edge modes remain illusive~\cite{Zhao2020,Zhao2023,Yuval2022,Hatefipour2022,uday2024}, due to presence of magnetic vortices, thermal fluctuations, etc. The propagating Bogoliubov edge modes give rise to flat density of states, which may be probed by scanning tunnelling spectroscopy~\cite{menard2017two,kezilebieke2020topological,wang2020evidence}. Therefore, it is highly desired to explore more spectroscopy signature of chiral Bogoliubov edge modes and TSC. 

\begin{figure}[t]
  \begin{center}
  \includegraphics[width=3.4in,clip=true]{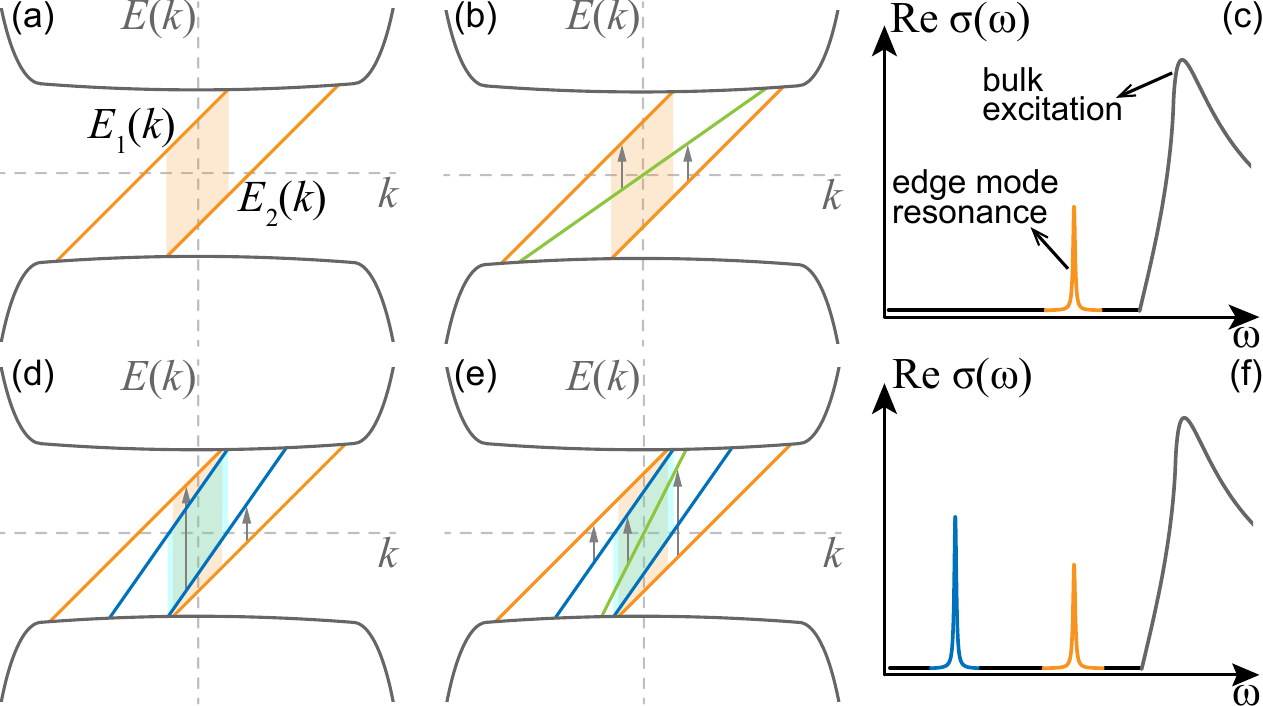}
  \end{center}
  \caption{Schematic diagrams for the energy dispersion and corresponding linear optical spectrum of chiral TSC with different BdG Chern numbers. (a) $N=2$ and (b) $N=3$ TSC have similar optical response in (c), with one resonance peak (orange line) below the frequency of bulk quasi-particle excitation (gray line), which is originated from the transition within the edge particle-hole pair bands (orange). (d) $N=4$ TSC has a similar shape of optical spectrum as (e) $N=5$, both of them have two pairs of Andreev edge modes (blue and orange), giving rise to two in-gap resonance peaks in (f) (blue and orange lines). For TSC with $N=3$ and $N=5$, the green lines represent the self-conjugate chiral Majorana modes. The optical transition between this mode and other modes does not alter the spectrum qualitatively. 
  The orange and blue shaded regions represent the momentum regions where intra-pair transitions occur, while inter-pair transitions are indicated by gray arrows.}
  \label{fig1}
\end{figure}

In this Letter, we study the optical effect of chiral TSC in two dimensions. We show that in the clean limit, where the normal state mean free path $\ell\gg\xi$ the coherent length of Cooper pairs, the linear optical response generically has in-gap resonances originated from the transitions within the particle-hole pair bands of chiral Bogoliubov edge modes [see Fig.~\ref{fig1}(a)]. Interestingly, the number of resonance peaks is related to the topological invariant of chiral TSC, which in turn serves as a spectroscopy method to probe TSC with different Chern numbers. The chiral Bogoliubov edge modes are \emph{on average} charge neutral, namely their electron and hole components have equal weight only on average. Hence if the two states involved in the optical transition have different charge distributions, a dipole oscillation is generated, enabling the response to the electromagnetic field~\cite{lu2022directly}. Quite different from the previous study on the local optical response of chiral edge modes in TSC~\cite{he2021opticala}, which requires a highly focused light beam on the edges involve only optical transitions within a single edge mode. We consider here the low frequency optical conductivity under uniform illumination, which is feasible in experiments, with the optical response arising from inter-edge-mode transitions.

\emph{Edge theory of optical response.}
We consider the uniform and normal incidence condition, where the momentum of photon is neglected. In the clean limit, only vertical optical transition is allowed. In this scenario, the general properties of the optical response of SC have been analyzed in Ref.~\cite{ahn2021theory}. For a single-band SC, if the normal state dispersion satisfies $\epsilon(\mathbf{k})=\epsilon(-\mathbf{k})$, which can be guaranteed by inversion symmetry or time-reversal symmetry, the current operator is zero and thus optical excitation is absent, due to $\mathfrak{C}$ symmetry, which combines particle-hole conjugation and a momentum-reversing unitary operation. An intuitive understanding of the absence of optical response is that when a Cooper pair, with zero total momentum, is broken by light, the two electrons with opposite momentum will have the same velocity but in opposite directions, resulting in zero net current~\cite{mahan2000,huang2023}. Since the current operators of the edge modes are obtained by projecting the bulk current operators to the edge states, the multiband condition is indispensable for a nonzero optical response of Bogoliubov edge modes. Chiral TSC from superconducting proximity of QAH state fullfils this condition, since a QAH insulator is naturally a multiband system because the band inversion always involve at least two bands with opposite parities. Without loss of generality, we will consider the optical effect of chiral TSC from superconducting proximity of QAH state for concreteness.

We first study the optical effect from the effective edge theory for the simplest case, where the QAH state has Chern number $C=1$. When the superconducting proximity gap is smaller than the the insulating bulk gap of the QAH state, the chiral TSC will be in the same topological phase as the QAH state of Chern number $C$ with charge $U(1)$ symmetry broken, which is a superconductor with Bogoliubov-de Gennes (BdG) Chern number $N=2C=2$~\cite{qi2010chiral,wang2015chiral}. The effective edge  Hamiltonian can be described as~\cite{lian2018}
\begin{equation}\label{ckH}
  \mathcal{H}_{\text{edge}}=\sum_{k>0}^{k_0}{\left(c^\dagger_k,c_{-k}\right)
  \begin{pmatrix}
    vk-\mu & -\Delta \\
    -\Delta & vk+\mu \\
  \end{pmatrix}
  \begin{pmatrix}
    c_k \\
    c^\dagger_{-k} \\
  \end{pmatrix}},
\end{equation}
where $c_k$ annihilates one electron with momentum $k$, $v$ is the velocity of the edge mode, $\mu$ is the chemical potential, $\Delta$ is the pairing gap function for the edge mode, and to the lowest order, $\Delta=\Delta_0 k$, $k_0$ is the momentum cutoff. We can decompose the electron operators into Majorana operators: $\gamma_{1,k}=(c^\dagger_{-k}+c_k)/\sqrt{2}$, $\gamma_{2,k}=i(c^\dagger_{-k}-c_k)/\sqrt{2}$, then the edge Hamiltonian becomes:
\begin{equation}\label{gkH}
  \mathcal{H}_{\text{edge}}=\sum_{k>0}^{k_0}{\left(\gamma^\dagger_{1,k},\gamma^\dagger_{2,k}\right)
  \begin{pmatrix}
    vk-\Delta & -i\mu \\
    i\mu & vk+\Delta \\
  \end{pmatrix}
  \begin{pmatrix}
    \gamma_{1,k} \\
    \gamma_{2,k} \\
  \end{pmatrix}}.
\end{equation}
The edge spectrum is $E_{1,2}(k)=vk\pm\sqrt{\Delta^2+\mu^2}$. The single chiral fermion mode of QAH state now splits into a particle-hole conjugate pair of chiral Bogoliubov mode of chiral TSC, satisfying $E_1(k)=-E_{2}(-k)$. They are also called chiral Andreev edge modes~\cite{Zhao2020}. The number of chiral edge modes is exactly equal to the bulk BdG Chern number from the bulk-edge correspondence. Generically, these two modes have a momentum difference as shown in Fig.~\ref{fig1}(a). Keeping up to the first order of spatial derivative, the current operator~\cite{bi2024vertical} along the direction of the boundary only takes the form as $j=\sum_{k>0}^{k_0}(ak\tau_0+b\tau_2)$, where $a$ and $b$ are real denoting the intrabranch and interbranch contribution, respectively, $\tau_i$ is the Pauli matrix in the basis of Eq.~(\ref{gkH}). Typically, the pairing amplitude is much smaller than the chemical potential. In this case, we show that in the clean limit, the optical absorption, namely the real part of the linear optical conductivity contains a single resonant frequency as~\cite{sm},
\begin{equation}\label{sigmaxx}
  \text{Re}\left[\sigma(\omega)\right] \sim \delta\left(\omega-2|\mu|\right).
\end{equation}

Here the response current is along the direction of edge modes. Such photoexcitation of chiral Bogoliubov edge modes can also be understood intuitively as follows: since 
the normal state dispersion satisfies: $\epsilon(k)=-\epsilon(-k)$, when a Cooper pair is broken by light into two Bogoliubov edge modes, they are charged and have the same velocity, leading to a finite net current. Generically the resonance frequency of two conjugate Bogoliubov edge modes lies below the bulk quasi-particle excitation frequency, thus the typical linear optical effect of the chiral TSC is shown in Fig.~\ref{fig1}(c). 

Next we consider general cases with $N>2$. For TSC with the BdG Chern number $N=2C$ ($C>1$), there will be $C$ particle-hole conjugate pairs of chiral Bogoliubov edge modes~\cite{qi2010chiral,wang2015chiral}. In general, these $C$ pairs of edge modes have different velocity and distinctly separated in momentum space, such as $N=4$ shown in Fig.~\ref{fig1}(d). For a given incident frequency, the resonant momentum region is determinted by energy conservation. For intra-pair transitions, the dispersion curves of two chiral Bogoliubov edge modes are nearly parallel, resulting in a relatively large resonant region. By contrast, for inter-pair transitions, their dispersion curves are generally not parallel, so resonances occur only at discrete momentum points, leading to a much smaller contribution. Thus intra-pair transitions exhibit distinct resonance peaks, whereas inter-pair transitions do not produce significant features in the optical spectrum. As a result, the whole optical spectrum should contain $C$ in-gap resonance peaks as illustrated in Fig.~\ref{fig1}(f) and calculated in Supplemental Material~\cite{sm} for $N=4$. For TSC with $N=2C+1$, there is one additional self-conjugate chiral Bogoliubov edge mode, also called chiral Majorana edge mode with dispersion $E(k)=v'k$~\cite{qi2010chiral}. Generically, the velocity $v'$ of this self-conjugate mode is different from other $C$ pairs, then the optical transition between it and other modes does not show resonance peaks, as it falls under the category of inter-pair transitions. Thus there are also $C$ in-gap resonance peaks as schematically shown in Fig.~\ref{fig1}(c) and Fig.~\ref{fig1}(f) for $N=3$ and $N=5$, respectively. In summary, TSC with $N=2C$ and $N=2C+1$ show similar edge optical responses, while TSC with different $C$ are expected to give rise to different edge responses. The different in-gap optical resonance structure could serve as a spectroscopy method to probe TSC with different Chern numbers, which distinguish integral part of $N/2$, namely $[N/2]$.

\emph{Bulk model calculation.}
Now we turn to optical response calculation of the bulk model. From the above analysis, chiral TSC with $N=1$ and $N=2$ exhibit different optical response features. As an example, we consider superconducting proximity of QAH state in a magnetic topological insulator thin film with ferromagnetic order~\cite{wang2015chiral}. The effective Hamiltonian for QAH state is $\mathcal{H}_0=\sum_{\mathbf{k}}\psi^{\dag}_{\mathbf{k}}H_0(\mathbf{k})\psi_{\mathbf{k}}$, with $\psi_{\mathbf{k}}=(c^t_{\mathbf{k}\uparrow}, c^t_{\mathbf{k}\downarrow},c^b_{\mathbf{k}\uparrow}, c^b_{\mathbf{k}\downarrow})^T$ and $H_0(\mathbf{k})=k_y\sigma_x\tau_z-k_x\sigma_y\tau_z+m(\mathbf{k})\tau_x+\lambda\sigma_z$. Here the superscripts $t$ and $b$ denote the top and bottom surface states, respectively. $\sigma_i$ and $\tau_i$ are Pauli matrices for spin and layer, respectively. $\lambda$ is the exchange field. $m(\mathbf{k})=m_0+m_1(k_x^2+k_y^2)$ is the hybridization between the top and bottom surface states. $\mathcal{H}_0$ well describes the QAH state in Cr-doped (Bi,Sb)$_2$Te$_3$~\cite{chang2013experimental} and odd layer MnBi$_2$Te$_4$~\cite{Zhang2019}. The BdG Hamiltonian for chiral TSC is $\mathcal{H}_{\mathrm{bulk}}=\sum_{\mathbf{k}}\Psi^{\dag}_{\mathbf{k}}H_{\mathrm{bulk}}(\mathbf{k})\Psi_{\mathbf{k}}/2$, with $\Psi_{\mathbf{k}}=[(c^t_{\mathbf{k}\uparrow}, c^t_{\mathbf{k}\downarrow}, c^b_{\mathbf{k}\uparrow}, c^b_{\mathbf{k}\downarrow}), (c^{t\dag}_{-{\mathbf{k}}\uparrow}, c^{t\dag}_{-{\mathbf{k}}\downarrow}, c^{b\dag}_{-{\mathbf{k}}\uparrow}, c^{b\dag}_{-{\mathbf{k}}\downarrow})]^T$ and
\begin{equation}\label{QAH+sc}
\begin{aligned}
  H_{\mathrm{bulk}}(\mathbf{k}) &=
    \begin{pmatrix}
      H_0(\mathbf{k})-\mu & \Delta(\mathbf{k})\\
      \Delta(\mathbf{k})^\dag & -H_0^*(-\mathbf{k})+\mu
    \end{pmatrix},\\
  \Delta(\mathbf{k}) &=
  \begin{pmatrix}
    i\Delta_t\sigma_y & 0\\
    0 & i\Delta_b\sigma_y
  \end{pmatrix}.
\end{aligned}
\end{equation}
Here $\mu$ is the chemical potential, $\Delta_t$ and $\Delta_b$ are pairing gap functions on top and bottom surface states, respectively. The topological properties of this system were well studied in Ref.~\cite{wang2015chiral}, revealing three TSC phases with BdG Chern number $N=0,1,2$. 

\begin{figure}[t]
  \begin{center}
  \includegraphics[width=3.4in,clip=true]{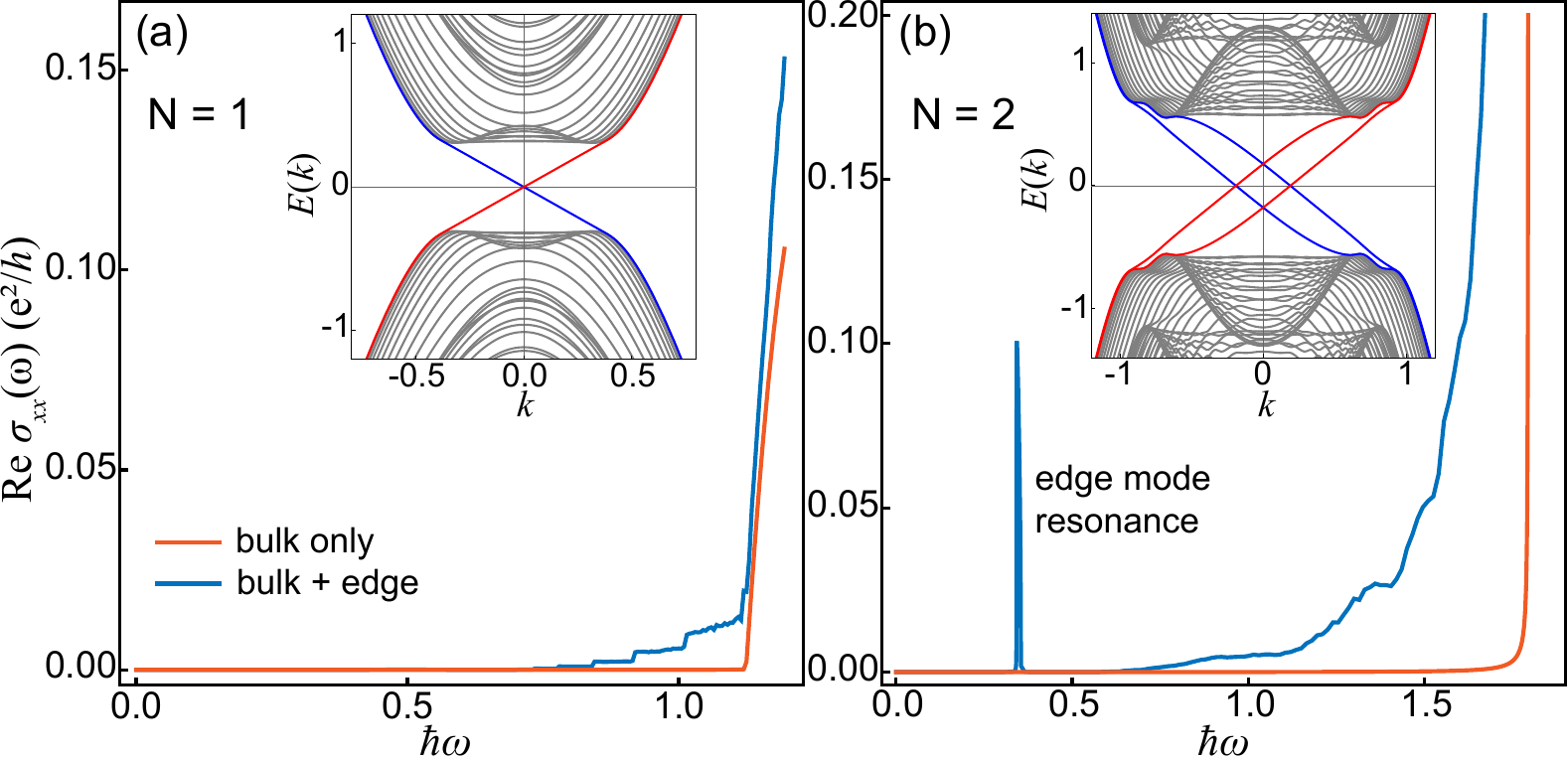}
  \end{center}
  \caption{(a), (b) $\text{Re}[\sigma(\omega)]$ of the chiral TSC phases with $N=1$ and $N=2$, respectively. The blue (orange) lines are calculated with open (periodic) boundary condition along $y$-direction, while $x$-direction is periodic, corresponding to the optical response from bulk bands with  (without) edge modes. The insets show the corresponding energy spectrum with open boundary along $y$-direction, where the blue and red chiral Bogoliubov edge modes are localized at opposite boundaries. The parameters are 
  $\mu=0.3$, $m_0=1$, $m_1=3$, $\Delta_t=1$, $\Delta_b=0$, $\eta=10^{-4}$. The system length $L_y=60$. We set $\lambda=1$ for the $N=1$ phase and $\lambda=3$ for the $N=2$ phase.}
  \label{fig2}
\end{figure}

The optical conductivity is calculated using the Kubo formula:
\begin{equation}\label{kubo}
  \sigma_{\alpha\beta}(\omega)= -\frac{i e^2}{\hbar}\sum_{k_x,m,n} \frac{f_{mn}(k_x)}{\epsilon_{mn}(k_x)}
  \frac{\langle j_\alpha\rangle_{mn}(k_x) \langle j_\beta\rangle_{nm}(k_x)}{\epsilon_{mn}(k_x)+\omega+i\eta},
\end{equation}
where $\eta$ is a infinitesimal positive parameter, $f_{mn}(k_x)\equiv f_m(k_x)-f_n(k_x)$, $f_n(k_x)$ is the Fermi-Dirac function of the eigenstate $|m(k_x)\rangle$, $\epsilon_m(k_x)$ is the corresponding eigenenergy, $\epsilon_{mn}(k_x)\equiv \epsilon_m(k_x)-\epsilon_n(k_x)$, $\alpha,\beta=x,y$, $\langle j_\alpha\rangle_{mn}(k_x)\equiv \langle m(k_x)|j_\alpha(k_x)|n(k_x) \rangle$ and $j_\alpha(k_x)$ is the velocity operator defined as:
\begin{equation}
  j_\alpha (k_x)=-
  \begin{pmatrix}
    \frac{\partial H_0(k_x)}{\partial k_\alpha} & 0 \\
    0 & \frac{\partial H^*_0(-k_x)}{\partial k_\alpha}
  \end{pmatrix}.
\end{equation}
In the configuration with periodic/open boundary condition along $x$/$y$-direction, the edge optical response leads to a resonance peak to $\text{Re}[\sigma_{xx}(\omega)]$ but does not contribute to $\text{Re}[\sigma_{yy}(\omega)]$, since the chiral Bogoliubov edge modes propagate along $x$-direction. 

The numerical results are shown in Fig.~\ref{fig2}. First, we note that the excitation of the lowest Bogoliubov bulk band is absent, and the threshold frequency corresponds to the transition from the lowest band to the second lowest band~\cite{lu2024optical}. This can be explained by the $\mathfrak{C}$ symmetry, which is the combination of particle-hole conjugation and twofold rotation $C_{2z}$. The bulk Hamiltonian satisfies $\mathfrak{C}{H}_{\text{bulk}}(\mathbf{k}) \mathfrak{C}^{-1}=-{H}_{\text{bulk}}(\mathbf{k})$, and $\mathfrak{C}^2=-1$. As proved in Ref.~\cite{ahn2021theory}, $\langle\mathfrak{C} \cdot n(\mathbf{k})|j_\alpha(\mathbf{k})|n(\mathbf{k})\rangle=0$, prohibiting the lowest-energy excitations.
While this symmetry is explicitly broken by the open boundary, the optical transition between chiral Bogoliubov edge modes can give rise to finite conductivity. Additionally, the transition between edge modes and bulk bands can also contribute to optical  conductivity, whose signal is close to the bulk excitation signal and does not resonate. 
Finally, we analyze the resonance feature of the spectrum. For the $N=1$ phase, there is only one chiral Majorana edge mode, which can only transit into the bulk band without any in-gap resonance peak. For the $N=2$ phase, there is a pair of chiral Andreev edge modes, which gives rise to one in-gap resonance peak. For the $N=0$ phase, no chiral Bogoliubov edge mode exist and hence no in-gap resonance peak in the optical absorption. Therefore, the absence or presence of the in-gap resonance peak is the main difference in the linear optical response between $N=0, 1$ and $N=2$ phases. This is consistent with the edge theory above that linear optical spectroscopy can only distinguish chiral TSC with the different integral parts of $N/2$.

\begin{figure}[t]
  \begin{center}
  \includegraphics[width=3.4in,clip=true]{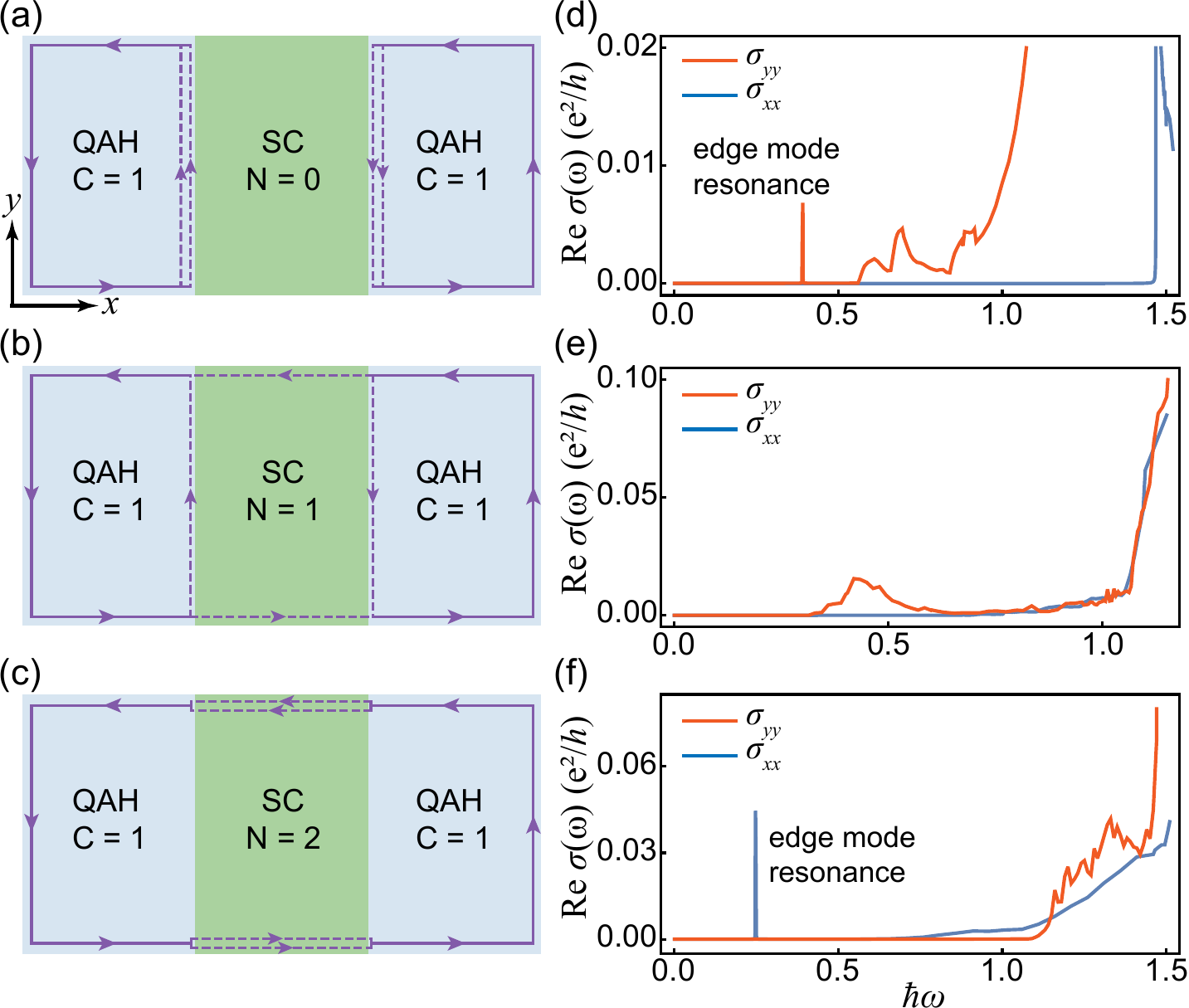}
  \end{center}
  \caption{The QAH-SC-QAH junction, where the BdG Chern number of the SC is $N=0, 1, 2$. The violet arrow lines at the interfaces represent chiral electron edge modes (solid lines) and chiral Bogoliubov edge modes (dashed lines). (d), (e), (f) are the corresponding optical responses for (a), (b), (c), respectively. We employ the model in Eq.~(\ref{QAH+sc}), and set $\mu=0.2, 0.2, 0.2$, $\lambda=0.3, 1, 3$, $\Delta_1=0.4, 1, 1$ for TSC with $N=0, 1, 2$, respectively, and $\mu=0.2$, $\lambda=3$, $\Delta_1=0$ for the QAH insulator, other parameters are the same in Fig.~\ref{fig2}.}
  \label{fig3}
\end{figure}

\emph{QAH-SC junction device.}
The optical response of $N=0$ and $N=1$ chiral TSC show qualitatively similar feature without in-gap resonance peaks, however, we propose that the optical absorption could distinguish them in a junction device. 
To illustrate this, we consider a QAH-SC-QAH heterojunction
shown in Fig.~\ref{fig3}(a)-\ref{fig3}(c), which was previously proposed as a non-Abelian gate to study the chiral Majorana backscattering~\cite{wang2015chiral,lian2018topological}. The TSC phase can be realized by inducing superconducting proximity in the middle region of QAH insulator. The schematics of edge modes configuration in this junction are shown in Fig.~\ref{fig3}, where the different TSC phases can be achieved by tuning external magnetic or electric field. The chiral TSC in Fig.~\ref{fig3}(b) has $N=1$, then each edge between the chiral TSC and the vacuum or QAH hosts a chiral Majorana edge mode denoted as dahsed line. 

Then we study the optical effect in such junctions, where the optical field illuminates the whole superconducting region. The edge modes are localized along both $x$- and $y$-directions, generating optical responses in $\sigma_{xx}$ and $\sigma_{yy}$, respectively. $\sigma_{yy}$ is calculated in a cylindrical geometry that is translationally invariant in the $y$-direction while forming a junction of $C=1$ QAH insulator and TSC in the $x$-direction~\cite{lian2016edgestateinduced}. While $\sigma_{xx}$ is calculated by setting open/periodic boundary conditions in the $y/x$-direction. For comparison, a $C=1$ QAH insulator cannot produce any in-gap resonance peaks~\cite{sm}. The numerical results of $\text{Re}[\sigma_{xx}(\omega)]$ and $\text{Re}[\sigma_{yy}(\omega)]$ for three cases in Fig.~\ref{fig3}(a)-\ref{fig3}(c) are shown in Fig.~\ref{fig3}(d)-\ref{fig3}(f), respectively. The $N=0$ phase has one in-gap resonance peak only in $\text{Re}[\sigma_{yy}(\omega)]$, which originated from the chiral Andreev edge modes at the interface between $C=1$ QAH and $N=0$ SC. While the chiral Andreev edge modes between $N=2$ phase and vacuum contribute to the in-gap resonance peak only in $\text{Re}[\sigma_{xx}(\omega)]$. The $N=1$ phase does not show a resonance peak in Fig.~\ref{fig3}(e). Therefore, the optical effect provides an applicable method to detect the chiral Bogoliubov edge modes configuration and hence the topological phases of the junction devices.

\emph{Experimental feasibility.}
We discuss the experimental feasibility of the proposed effect. Firstly, we estimate the in-gap resonant frequency. A typical proximity-induced superconducting gap is on the order of $0.1$~meV~\cite{uday2024,wang2012coexistence}, and the energy separation of the chiral Bogoliubov edge modes is about one order of magnitude smaller, namely $0.01$~meV. This energy corresponds to a frequency $f\approx2.4$~GHz, which is in the microwave regime. The typical ferromagnetic resonance frequency of a magnetic topological insulator is approximately $10$~GHz~\cite{liu2020,liu2022dynamical}, which can be tuned by an external field, allowing it to be distinguished from the optical resonance frequency of edge modes.

Secondly, we have only considered the $\ell\gg\xi$ limit. When electrons are depaired by the electromagnetic field during a characteristic time scale $\ell/v_F$, where $v_F$ is the Fermi velocity of electrons in the normal state, they are nearly not scattered by disorder within a characteristic time scale $\xi/v_F$, and their momenta are conserved~\cite{xiang2022dwave}. MnBi$_2$Te$_4$ may fullfil this condition, where the mobility is about $2\times10^3$~cm$^{2}$/V$\cdot$s, Fermi velocity is $5.5\times10^5$~m/s, then $\ell\sim1$~$\mu$m~\cite{deng2020quantum}. The superconducting coherent length is of about 0.1~$\mu$m~\cite{uday2024}. The strong disorder scattering will change the momentum of electrons, so the optical transition would not be restricted to vertical transitions. Such processes will smear the in-gap resonance peak.

\emph{Discussion.}
We discuss the temperature dependence of the optical absorption of chiral Bogoliubov edge modes, where only $f_{mn}(k_x)$ in Eq.~(\ref{kubo}) has temperature dependence. This term is maximized when one mode is fully occupied and the other is completely unoccupied, corresponding to the zero-temperature limit. As the temperature increases (but still $\ll T_c$), the difference in Fermi distribution functions decreases, leading to a reduction and broadening of the resonance peak~\cite{sm}. 

Besides the quasiparticle excitation of bulk or edge states, various collective modes could contribute to the optical absorption, such as phase modes~\cite{anderson1958randomphasea,anderson1958new,bogoljubov1958new,rickayzen1959collective}, amplitude modes~\cite{littlewood1982amplitude,pekker2015amplitude,shimano2020higgs}, etc. 
However, the collective modes usually do not have linear coupling to the electromagnetic field and hence do not exhibit a linear response.
Furthermore, some types of collective modes exist only in SC with richly structured order parameters, which is not the case considered here.
Therefore, the in-gap resonance in linear optical spectroscopy only involves quasiparticle excitation and is attributed to the transition between chiral Bogliubov edge modes.

Finally, we discuss the QAH-SC junction device with ineffective SC proximity.
One possible scenario is that the whole device is equivalent to a $C=1$ QAH system, where the optical response of this system has no in-gap resonance~\cite{sm}. An alternate scenario is the middle region becomes non-superconducting metallic puddles due to inhomogeneity. The metallic puddles are gapless in the bulk, resulting in a strong signal in $\text{Re}[\sigma(\omega)]$. This signal is Drude-like, peaking at zero frequency and decaying as a power law with increasing frequency, which is quite different from the chiral Bogoliubov edge modes with a finite frequency resonance peak.

The results can be directly extended to the QH state with SC proximity. The QH edge states are degenerate, thus there is no finite-frequency optical resonance between them. Moreover, the energy difference between Landau levels is typically much larger than the superconducting gap, so their resonance frequency cannot be in-gap. In addition, a SC has divergent imaginary conductivity $\text{Im}[\sigma(\omega)]$ at low frequencies, scaling as $1/\omega$, reflecting the Meissner effect~\cite{tinkham2004introduction}, while a metal or a dielectric medium has vanishing imaginary conductivity as $\omega \rightarrow 0$. Therefore, these phases exhibit different spectral behaviors and can be distinguished.

In summary, the optical response within the particle-hole conjugate Bogoliubov edge modes leads to resonance peak below the bulk quasiparticle excitation frequency, which provides a new applicable method for detecting chiral Bogoliubov edge states and is distinguishable from collective modes in SC and possible trivial explanations. Optical spectroscopy offers a valuable toolbox for exploring novel topological SC.

\begin{acknowledgments}
\emph{Acknowledgment.} This work is supported by the National Key Research Program of China under Grant No.~2019YFA0308404, the Natural Science Foundation of China through Grants No.~12350404 and No.~12174066, the Innovation Program for Quantum Science and Technology through Grant No.~2021ZD0302600, the Science and Technology Commission of Shanghai Municipality under Grants No.~23JC1400600, No.~24LZ1400100 and No.~2019SHZDZX01.
\end{acknowledgments}

\pagebreak
\widetext
\clearpage
\begin{center}
\textbf{\large Supplementary Material for ``Edge Optical Effect as Probe of Chiral Topological Superconductor"}
\end{center}
%%%%%%%%%% Prefix a "S" to all equations, figures, tables and reset the counter %%%%%%%%%%
\setcounter{equation}{0}
\setcounter{figure}{0}
\setcounter{table}{0}
\setcounter{page}{1}
\makeatletter
\renewcommand{\theequation}{S\arabic{equation}}
\renewcommand{\thefigure}{S\arabic{figure}}
\renewcommand{\thesection}{\Roman{section}}

\section{I. Derivation of Eq.~(\ref{sigmaxx}) in the main text}\label{app 1}
In this section, we provide the derivation of the optical response for the transition between two chiral Bogoliubov edge modes. 
We start from the edge Hamiltonian, Eq.~(\ref{gkH}):
\begin{equation}\label{gkH2}
  H=\sum_{0<k<k_0}{\left(\gamma^\dagger_{1,k}~\gamma^\dagger_{2,k}\right)
  \begin{pmatrix}
    vk-\Delta & -i\mu \\
    i\mu & vk+\Delta \\
  \end{pmatrix}
  \begin{pmatrix}
    \gamma_{1,k} \\
    \gamma_{2,k} \\
  \end{pmatrix}}.
\end{equation}
Since the two edge modes are coupled, the corresponding current operator also includes the coupling between these two modes. Up to the first order of spatial derivative and the coupling between $\gamma_{1,k}$ and $\gamma_{2,k}$, the current operator takes the form \cite{bi2024vertical}:
\begin{equation}
  \begin{aligned}
      j=\sum_{0<k<k_0} \big[ ak(\gamma^\dagger_{1,k}\gamma_{1,k}+\gamma^\dagger_{2,k}\gamma_{2,k})+ ib(-\gamma^\dagger_{1,k}\gamma_{2,k}+\gamma^\dagger_{2,k}\gamma_{1,k})\big]
       =\sum_{0<k<k_0}(ak\tau_0+b\tau_2).
    \end{aligned}
\end{equation}
To find the eigenstate representation, the Hamiltonian Eq.~(\ref{gkH2}) can be diagonalized as:
\begin{equation}\label{gtkH}
  H=\sum_{0<k<k_0}{\left(\tilde{\gamma}^\dagger_{1,k}~\tilde{\gamma}^\dagger_{2,k}\right)
  \begin{pmatrix}
    E_{1,k} & 0 \\
    0 & E_{2,k} \\
  \end{pmatrix}
  \begin{pmatrix}
    \tilde{\gamma}_{1,k} \\
    \tilde{\gamma}_{2,k} \\
  \end{pmatrix}},
\end{equation}
under the following unitary transformation:
\begin{equation}\label{trans}
  \begin{pmatrix}
    \tilde{\gamma}_{1,k} \\
    \tilde{\gamma}_{2,k} \\
  \end{pmatrix}
  =
  \begin{pmatrix}
    u_k & v_k \\
    -v^*_k & u^*_k \\
  \end{pmatrix}
  \begin{pmatrix}
    \gamma_{1,k} \\
    \gamma_{2,k} \\
  \end{pmatrix},
\end{equation}
where $E_{1,k}=vk-\sqrt{\Delta^2+\mu^2},~E_{2,k}=vk+\sqrt{\Delta^2+\mu^2}$ are the eigenenergies. The $u_k$ and $v_k$ are chosen as:
\begin{equation}
  \begin{aligned}
    u_k=i u=-i\sqrt{\frac{1}{2}+\frac{\Delta}{2\sqrt{\Delta^2+\mu^2}}},\qquad
    v_k=v=\sqrt{\frac{1}{2}-\frac{\Delta}{2\sqrt{\Delta^2+\mu^2}}}.
  \end{aligned}
\end{equation}
In this basis, the current operator can be divided into diagonal and off-diagonal parts as:
\begin{equation}
  \begin{aligned}
    \tilde{j}=\sum_{0<k<k_0}&\tilde{j}_{1,k}+\tilde{j}_{2,k}, \\
    \tilde{j}_{1,k}=ak\tau_0+bi(v_k u^*_k-u_k v^*_k)\tau_3,\qquad
    &\tilde{j}_{2,k}=bi
    \begin{bmatrix}
      0 & -(u^2_k+v^2_k) \\
      ({u^*_k})^2+({v^*_k})^2 & 0 \\
    \end{bmatrix}.
  \end{aligned}
\end{equation}

The real part of the optical conductivity is related to the imaginary part of the current-current correlation function $\Pi_{\alpha\beta}(\omega)$:
\begin{equation}
  \text{Re}~\sigma_{\alpha\beta}(\omega)=-\frac{1}{\omega V}\text{Im}~\Pi_{\alpha\beta}(\omega).
\end{equation}
where $V$ is the volume of the system. For a one-dimensional system, $V$ is the length of the system $L$.
In the following we only considered the longitudinal part along the direction of edge
modes, $\sigma_{xx}(\omega)$, thus $\alpha=\beta=x$ and will be omitted from now on.
The current-current correlation function $\Pi(\omega)$ can be obtained by analytic continuation: $\Pi(\omega)=\Pi(i \omega_n\rightarrow\omega+i\eta)$ with $\eta$ being an infinitesimal positive parameter, and
\begin{equation}
  \Pi(i \omega_n)=-\int_{0}^{\beta} \mathrm{d}\tau e^{i\omega_n\tau} T_\tau\langle \tilde{j}^\dagger(\tau)\tilde{j}(0) \rangle.
\end{equation}
$\Pi(i \omega_n)$ has two types of contributions: one contains the same current operator, denoted as $\Pi_{11}$ and $\Pi_{22}$, and the other is the cross term of $\tilde{j}_1$ and $\tilde{j}_2$, $\Pi_{12}$ and $\Pi_{21}$. Since the latter involves three $\tilde{\gamma}_{1(2)}$ and one $\tilde{\gamma}_{2(1)}$ operator, it vanishes due to the diagonal nature of the Green's function.
The correlation function of two $\tilde{j}_1$ operators can be shown to be zero:
\begin{equation}
  \begin{aligned}
  \Pi_{11}(i \omega_n)&=-\sum_{k,k'}\sum_{l=1,2}\int_{0}^{\beta} \mathrm{d}\tau e^{i\omega_n \tau} T_\tau\langle \tilde{\gamma}^\dagger_{l,k}(\tau) \tilde{\gamma}_{l,k}(\tau) \tilde{\gamma}^\dagger_{l,k'}(0) \tilde{\gamma}_{l,k'}(0) \rangle \\
  & = -\frac{1}{\beta} \sum_{k}\sum_{l=1,2}\sum_{\nu_n}{G_l(k,i\nu_n) G_l(k,i\omega_n+i\nu_n)}\\
  & = -\sum_{k}\sum_{l=1,2}\frac{f_{l,k}-f_{l,k}}{i\omega_n+E_{l,k}-E_{l,k}} \\
  & = 0,
  \end{aligned}
\end{equation}
where $f_{l,k}=f(E_{l,k})$ is the Fermi-Dirac distribution function, $\beta=1/k_B T$, $\nu_n=(2n+1)\pi/\beta$ is the Matsubara frequency. This result can be understood intuitively:
the photon momentum can be neglected, thus in the clean limit, only the vertical transition is allowed, prohibiting intra-mode response.
This also demonstrates the absence of edge optical response in the $N=1$ TSC, since there is only one Bogoliubov mode.
For $\Pi_{22}(i \omega_n)$, it can be found that:
\begin{equation}
  \begin{aligned}
    \Pi_{22}(i \omega_n)&=\frac{1}{\beta}\sum_{\nu_n}\sum_{0<k<k_0} b^2(1-4u^2v^2)[G_1(k,i\omega_n)G_2(k,i\omega_n+i\nu_n) + G_2(k,i\omega_n)G_1(k,i\omega_n+i\nu_n)] \\
    &= b^2 \sum_{0<k<k_0} \frac{\Delta^2}{\Delta^2+\mu^2} \left[\frac{f_{1,k}-f_{2,k}}{i\omega_n+E_{1,k}-E_{2,k}} + \frac{f_{2,k}-f_{1,k}}{i\omega_n+E_{2,k}-E_{1,k}}\right],
  \end{aligned}
\end{equation}
thus after analytic continuation, the current-current correlation function reads:
\begin{equation}\label{Pixx}
    \Pi(\omega)=\Pi_{11}(\omega)+\Pi_{22}(\omega)=b^2\sum_{0<k<k_0}\frac{\Delta^2}{\Delta^2+\mu^2}\bigg(\frac{f_{1,k}-f_{2,k}}{\omega-2\sqrt{\Delta^2+\mu^2}+i\eta}+\frac{f_{2,k}-f_{1,k}}{\omega+2\sqrt{\Delta^2+\mu^2}+i\eta}\bigg).
\end{equation}
To the lowest order, the pairing term is given by $\Delta = \Delta_0 k$. Taking the imaginary part in the clean limit, Eq.~(\ref{Pixx}) yields two Dirac delta functions at the resonance frequencies. For $\omega>0$, only the first term contributes to the real part of the optical conductivity. Typically, the pairing amplitude satisfies $\Delta_0 \ll \mu$, allowing the denominator of the first term to be approximated as $\omega-2|\mu|+i\eta$. Thus, the real part of the conductivity takes the form:
\begin{equation}\label{Ssigmaxx}
  \text{Re}~\sigma_{xx}(\omega) \sim \delta(\omega-2|\mu|).
\end{equation}
which is just Eq.~(\ref{sigmaxx}) in the main text. $\text{Re}~\sigma_{xx}(\omega)$ exhibits a resonance peak at about $\omega=2|\mu|$, corresponding to the optical transition between the two modes. In Fig.~(\ref{fig4})(a), we present the numerical results with parameters set as $v=1$, $\Delta_0=0.1$, $\mu=2$, $a=1$, $b=1$, and $\eta=10^{-4}$. The resonance peak is clearly observed at $\omega \sim 2\mu$, consistent with Eq.~(\ref{Ssigmaxx}).

\section{II. Optical response of two pairs of chiral Bogoliubov edge modes}
In this section, we calculate the optical spectrum of two pairs of chiral Bogliubov edge modes in a chiral topological superconductor with a Bogliubov-de Gennes Chern number $N=4$.
We start with the effective edge Hamiltonian of the quantum anomalous Hall insulator with Chern number $C=2$ and induced superconducting pairing by the proximity effect.
The effective edge Hamiltonian is:

\begin{equation}
  H=\sum_{0<k<k_0}{\Psi^\dagger_k
  \begin{pmatrix}
    v_1 k-\mu_1 & \lambda & -\Delta_1 & 0 \\
    \lambda & v_2 k-\mu_2 & 0 & -\Delta_2 \\
    -\Delta_1 & 0 & v_1 k+\mu_1 & -\lambda \\
    0 & -\Delta_2 & -\lambda & v_2 k+\mu_2 \\
  \end{pmatrix}
  \Psi_k},
\end{equation}
where $\Psi_k=(c_{1,k},c_{2,k},c^\dagger_{1,-k},c^\dagger_{2,-k})^T$, $\Delta_1$ and $\Delta_2$ are SC pairing functions, $\lambda$ is the coupling between two edge modes, $v_1$ and $v_2$ are velocities, $\mu_1$ and $\mu_2$ are chemical potentials.
In the Majorana basis: $\gamma_{1,k}=(c^\dagger_{1,-k}+c_{1,k})/\sqrt{2}$, $\gamma_{2,k}=i(c^\dagger_{1,-k}-c_{1,k})/\sqrt{2}$, $\gamma_{3,k}=(c^\dagger_{2,-k}+c_{2,k})/\sqrt{2}$, $\gamma_{4,k}=i(c^\dagger_{2,-k}-c_{2,k})/\sqrt{2}$, the Hamiltonian becomes:

\begin{equation}\label{Hgk2}
  H=\sum_{0<k<k_0}{\Gamma^\dagger_k
  \begin{pmatrix}
    v_1 k-\Delta_1 & -i \mu_1 & 0 & i \lambda \\
    i \mu_1 & v_1 k+\Delta_1 & -i \lambda & 0 \\
    0 & i \lambda & v_2 k-\Delta_2 & -i \mu_2 \\
    -i \lambda & 0 & i \mu_2 & v_2 k+\Delta_2 \\
  \end{pmatrix}
  \Gamma_k},
\end{equation}
where $\Gamma_k=(\gamma_{1,k},\gamma_{2,k},\gamma_{3,k},\gamma_{4,k})^T$. For simplicity, we consider $v_1=v_2=v>0$, $\mu_1=\mu_2=\mu>\lambda>0$, $\Delta>0$.
This Hamiltonian can be diagonalized as: $H=\sum_{0<k<k_0}{\tilde{\Gamma}^\dagger_k \text{diag} \{E_{1,k},E_{2,k},E_{3,k},E_{4,k}\} \tilde{\Gamma}_k}$,
where $E_{1,k}=v k -\sqrt{\Delta^2+(\lambda+\mu)^2}$, $E_{2,k}=v k -\sqrt{\Delta^2+(\lambda-\mu)^2}$, $E_{3,k}=v k + \sqrt{\Delta^2+(\lambda-\mu)^2}$, $E_{4,k}=v k +\sqrt{\Delta^2+(\lambda+\mu)^2}$. $\tilde{\Gamma}_k= U \Gamma_k$ is the unitary transformation, and $U$ can be chosen as:
\begin{equation}
  \begin{pmatrix}
    -i s_1 & i s_2 & -i t_2 & i t_1  \\
    -t_1   & t_2   & s_2    & -s_1   \\
    i s_1  & i s_2 & -i t_2 & -i t_1 \\
    t_1    & t_2   & s_2    & s_1    \\
  \end{pmatrix},
\end{equation}
with
\begin{equation}
  s_1,t_1=\frac{1}{2}\sqrt{1\pm\frac{\Delta}{\sqrt{\Delta^2+(\lambda+\mu)^2}}},  \qquad
  s_2,t_2=\frac{1}{2}\sqrt{1\pm\frac{\Delta}{\sqrt{\Delta^2+(\lambda-\mu)^2}}}.
\end{equation}

Due to the form of the Hamiltonian, Eq.~(\ref{Hgk2}), the current operator takes the form:
\begin{equation}
  j=\begin{pmatrix}
     a k & -i b & 0 & i c \\
     i b & a k & -i c & 0 \\
     0 & i c & a k & -i b \\
     -i c & 0 & i b & a k \\
    \end{pmatrix},
\end{equation} 
and in the eigenstate basis:
\begin{equation}
  \tilde{j}=\begin{pmatrix}
    a k - 4(b+c)s_1 t_1 & 0 & 0 & -2(b+c)(s^2_1-t^2_1) \\
    0 & a k - 4(b-c)s_2 t_2 & -2(b-c)(s^2_2-t^2_2) & 0 \\
    0 & -2(b-c)(s^2_2-t^2_2) & a k + 4(b-c)s_2 t_2 & 0 \\
    -2(b+c)(s^2_1-t^2_1) & 0 & 0 & a k + 4(b+c)s_1 t_1 \\
   \end{pmatrix}.
\end{equation}
Again, the current operator can be divided into diagonal and off-diagonal parts, and $\Pi_{11}=\Pi_{12}=\Pi_{21}=0$, only the intra-pair transition gives a nonzero contribution:
\begin{equation}
  \begin{aligned}
    \Pi_{22}(i \omega_n) & =\frac{1}{\beta}\sum_{\nu_n}\sum_{0<k<k_0} \bigl\{(b+c)^2(s^2_1-t^2_1)^2[G_1(k,i\omega_n)G_4(k,i\omega_n+i\nu_n) + G_4(k,i\omega_n)G_1(k,i\omega_n+i\nu_n)] \\
    & \qquad\qquad\qquad\quad +4(b-c)^2(s^2_2-t^2_2)^2[G_2(k,i\omega_n)G_3(k,i\omega_n+i\nu_n) + G_3(k,i\omega_n)G_2(k,i\omega_n+i\nu_n)] \bigr\} \\
    & =\sum_{0<k<k_0} \frac{(b+c)^2 \Delta^2}{\Delta^2+(\lambda+\mu)^2} \left[ \frac{f_{14}(k)}{i\omega_n+E_{14}(k)} + (1 \leftrightarrow 4) \right] +
    \frac{(b-c)^2 \Delta^2}{\Delta^2+(\lambda-\mu)^2} \left[ \frac{f_{23}(k)}{i\omega_n+E_{23}(k)} + (2 \leftrightarrow 3) \right],
  \end{aligned}
\end{equation}
where $f_{ij}(k)=f_{i,k}-f_{j,k},~E_{ij}(k)=E_{i,k}-E_{j,k}$ and $i,j=1,2,3,4$. Therefore, when $\omega
>0$:

\begin{equation}
  \begin{aligned}
  \text{Re}~\sigma_{xx}(\omega) & =-\frac{1}{\omega L}\text{Im}~\Pi_{\alpha\beta}(\omega) \\
  & =\frac{1}{2 \omega} \left[ \int_0^{k_0} \frac{(b+c)^2 \Delta^2}{\Delta^2+(\lambda+\mu)^2} {f_{14}(k)\delta[\omega-E_{41}(k)]} \mathrm{d}k + \int_0^{k_0} \frac{(b-c)^2 \Delta^2}{\Delta^2+(\lambda-\mu)^2} {f_{23}(k)\delta[\omega-E_{32}(k)]}\mathrm{d}k \right] \\
  & \sim \frac{1}{2 \omega} \left[ \int_0^{k_0} \frac{(b+c)^2 \Delta^2}{\Delta^2+(\lambda+\mu)^2} {f_{14}(k)} \mathrm{d}k \delta\left(\omega-2|\mu+\lambda|\right)+ \int_0^{k_0} \frac{(b-c)^2 \Delta^2}{\Delta^2+(\lambda-\mu)^2} {f_{23}(k)}\mathrm{d}k \delta\left(\omega-2|\mu-\lambda|\right) \right] .
  \end{aligned}
\end{equation}
In the third line, we have assumed $\Delta=\Delta_0 k$, and $\Delta_0 k_0, \lambda \ll \mu$. Clearly, the real conductivity shows two resonance peaks at approximately $\omega=2|\mu+\lambda|$ and $2|\mu-\lambda|$, which come from the two intra-pair transitions.

\begin{figure}[htbp]
  \begin{center}
  \includegraphics[width=6.8in,clip=true]{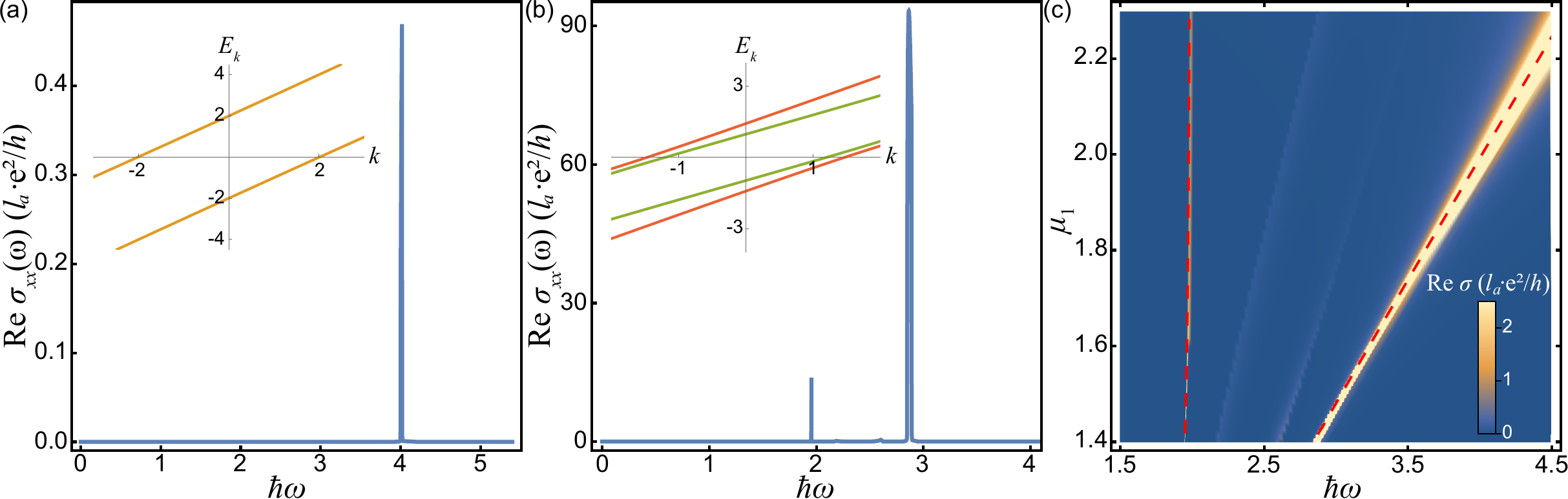}
  \end{center}
  \caption{Real conductivity for one pair (a) and two pairs (b) of chiral Bogliubov edge modes. The insets show the dispersions of edge modes. (c) Real conductivity for two pairs of modes as functions of $\mu_1$ and $\hbar\omega$. Since we consider a 1D edge model, the unit of 1D conductivity, compared to the 2D conductivity, includes the lattice constant $l_a$.}
  \label{fig4}
\end{figure}

In Fig.~(\ref{fig4})(b), we numerically calculate the real conductivity for a more general parameter setting: $v_1=1$, $\mu_1=2.5$, $v_2=0.8$, $\Delta_1=\Delta_2=0.1$, $\mu_2=1$, $\lambda=0.1$, $a_1=v_1$, $a_2=v_2$, $b_1=\mu_1$, $b_2=\mu_2$, $c=\lambda$, and $\eta=10^{-4}$. Then we vary $\mu_1$ from 1.4 to 2.3 in Fig.~(\ref{fig4})(c). From Fig.~(\ref{fig4})(b) and Fig.~(\ref{fig4})(c), we observe two prominent peaks at the frequencies of intra-pair mode resonance and some small values at other frequencies, corresponding to inter-pair transitions.
In Fig.~(\ref{fig4})(c), we mark $\sqrt{4\lambda^2+(\mu_1-\mu_2)^2}\pm(\mu_1+\mu_2)$ (red dashed lines), which are the energy differences between the modes in the two pairs and match well with the two resonance peaks. These results support the analysis in the main text.

\section{III. Optical response of a $C=1$ QAH insulator}\label{app 3}
\begin{figure}[htbp]
  \begin{center}
  \includegraphics[width=3.0in,clip=true]{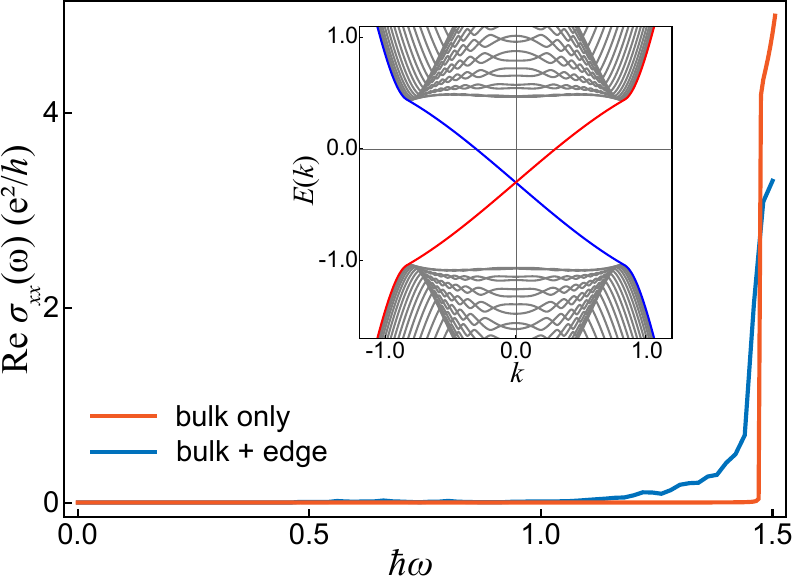}
  \end{center}
  \caption{The real part of the optical conductivity of the $C=1$ QAH. Insert shows the corresponding energy spectrum.}
  \label{fig5}
\end{figure}

In this section, we examine the optical response of the edge modes in a $C=1$ QAH system. As mentioned in the main text, if the superconducting proximity is not efficient in the QAH-SC hybrid device, only the chiral fermion mode of the QAH exists at the boundary, making the optical response of this device equivalent to that of a $C=1$ QAH.
In the numerical calculation, we use the model from Eq.~(\ref{QAH+sc}) in the main text and set $\Delta_1=0$ while leaving other parameters unchanged.
As shown in Fig.~\ref{fig5}, because there is only one chiral edge mode on one side, it is expected that no in-gap optical transition will occur. This is different from an $N=2$ TSC, whose optical spectrum is shown in Fig.~\ref{fig2}(b) in the main text, even if they are topologically equivalent.

\section{IV. Temperature effect on the edge resonance peak}\label{app 4}
\begin{figure}[htbp]
  \begin{center}
  \includegraphics[width=5in,clip=true]{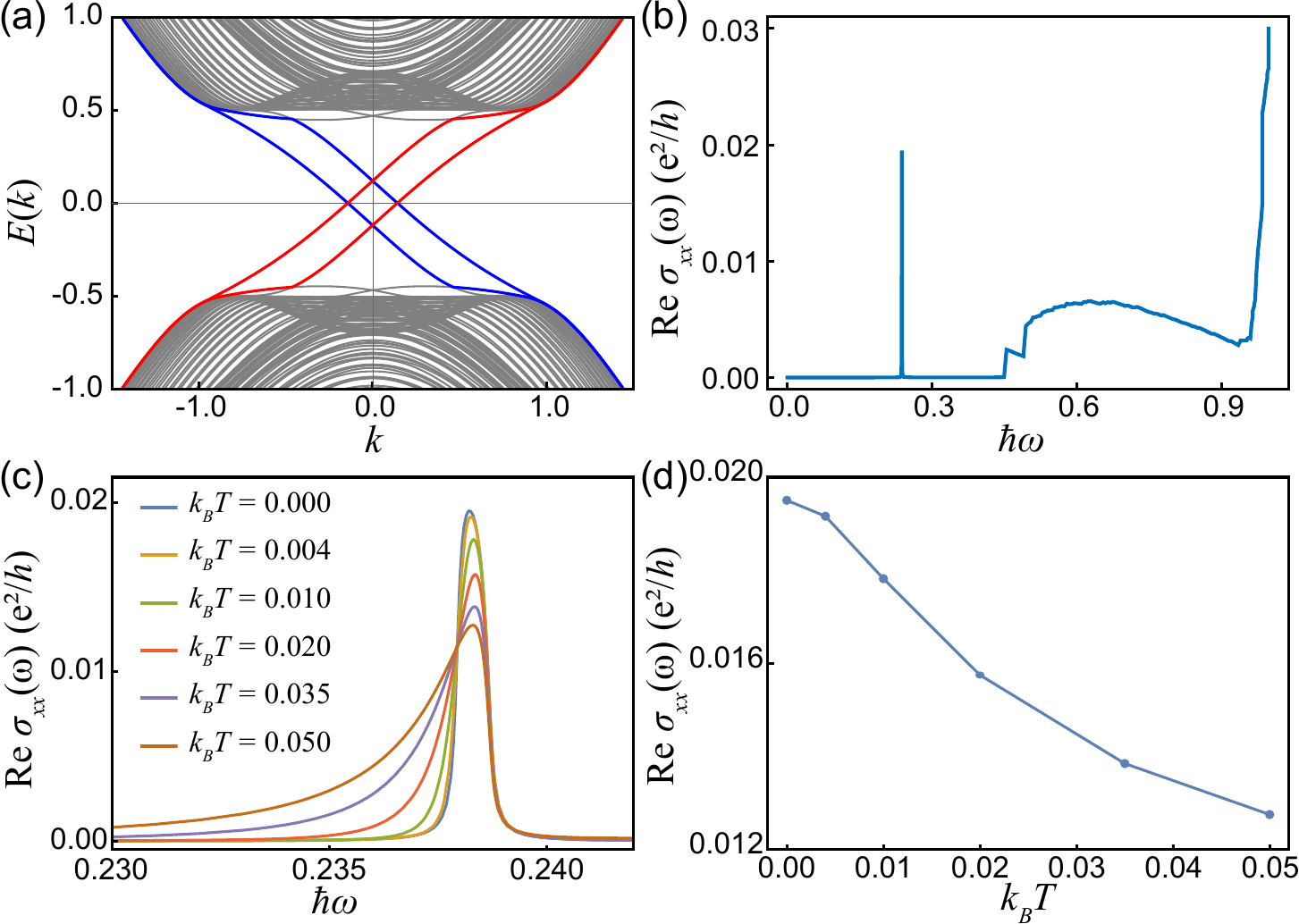}
  \end{center}
  \caption{(a) The temperature dependence of the optical response of the edge modes. (b) The temperature dependence of the peak value of the resonance peak.}
  \label{fig6}
\end{figure}

In this section, we present the temperature effect on the optical resonance peak of the optical transition of the chiral edge modes in the low temperature region ($T \ll T_c$).
We consider a QAH-NSC junction, which is translationally invariant in the $y$-direction while forming a junction of a $C=1$ QAH insulator and an $s$-wave superconductor (NSC) in the $x$-direction \cite{lian2016edgestateinduced}.
The QAH insulator can be described by:
\begin{equation}\label{HQAH}
  \mathcal{H}_{\text{QAH}}=\sum_{\mathbf{k}}c^\dag_\mathbf{k}\left[\bm{\zeta}(\mathbf{k})\cdot\bm{\sigma}-\mu_{h}\right]c_{\mathbf{k}},
\end{equation}
and the BdG Hamiltonian for the NSC is:
\begin{equation}\label{HNSC}
  \mathcal{H}_{\text{NSC}}=\sum_{\mathbf{k}}\Psi^\dagger_{\mathbf{k}}\left[\epsilon(\mathbf{k})-\mu_{s}\right]\Psi_{\mathbf{k}}+(\Delta \Psi_{\mathbf{k}}^T i\sigma_y \Psi_{-\mathbf{k}}+\mathrm{H.c.}).
\end{equation}
where $\Psi_{\mathbf{k}}=(c_{\mathbf{k}\uparrow},c_{\mathbf{k}\downarrow})^T$, $\bm{\zeta}(\mathbf{k})=[M-B(\cos k_xa+\cos k_ya), A\sin k_x a, A\sin k_y a ]$, $\bm{\sigma}=(\sigma_x,\sigma_y,\sigma_z)$ are the Pauli matrices, $\epsilon(\mathbf{k})=B(2-\cos k_xa-\cos k_ya)$ is the kinetic energy, $a$ is the lattice constant, $\Delta$ is the pairing amplitude, and $\mu_h$ and $\mu_s$ are the chemical potentials of the QAH insulator and the NSC, respectively.
The energy spectrum is shown in Fig.~\ref{fig6}(a), where the parameters are chosen as: $a=0.8$, $B=1.5625$, $M=2.625$, $A=1.25$, $\Delta=0.5$, $\mu_h=0$ and $\mu_s=0.5$. Along the $x$-direction, both the numbers of sites for the QAH insulator and NSC are set as 50.
Fig.~\ref{fig6}(b) shows the optical response of this system at zero temperature, where the in-gap resonance is contributed by the resonance between two chiral edge modes at the boundary between the QAH insulator and NSC.

Then we focus on the region near the in-gap resonance frequency and consider temperatures up to $k_B T=0.05$. In this temperature region, the gap function is nearly unchanged, and we only consider the effect of temperature on the Fermi-Dirac distribution function. The results are shown in Fig.~\ref{fig6}(c)(d). 
As the temperature increases, the resonance peak decreases, which is simply due to the decrease in the difference of the Fermi-Dirac functions of the two edge modes. Above the resonance frequency, all the response functions exhibit similar attenuation behavior, reflecting the Lorentzian shape, which can be captured by the effective edge theory. However, below the resonance frequency, the decay of the response function slows down as the temperature increases. This results from the specific dispersion of the edge modes.
As can be seen in Fig.~\ref{fig6}(a), the dispersion of the edge modes bends when the momentum is large, i.e., when the energy is close to the bulk, leading to the smallness of the energy difference between the two chiral modes. When the temperature increases, optical transitions can occur in this momentum region, which can contribute to the optical conductivity. Therefore, the decay of the response function deviates from the Lorentz line shape.

\section{V. Double $d+id$-wave superconducting system}\label{app 5}
\begin{figure}[htbp]
  \begin{center}
  \includegraphics[width=5in,clip=true]{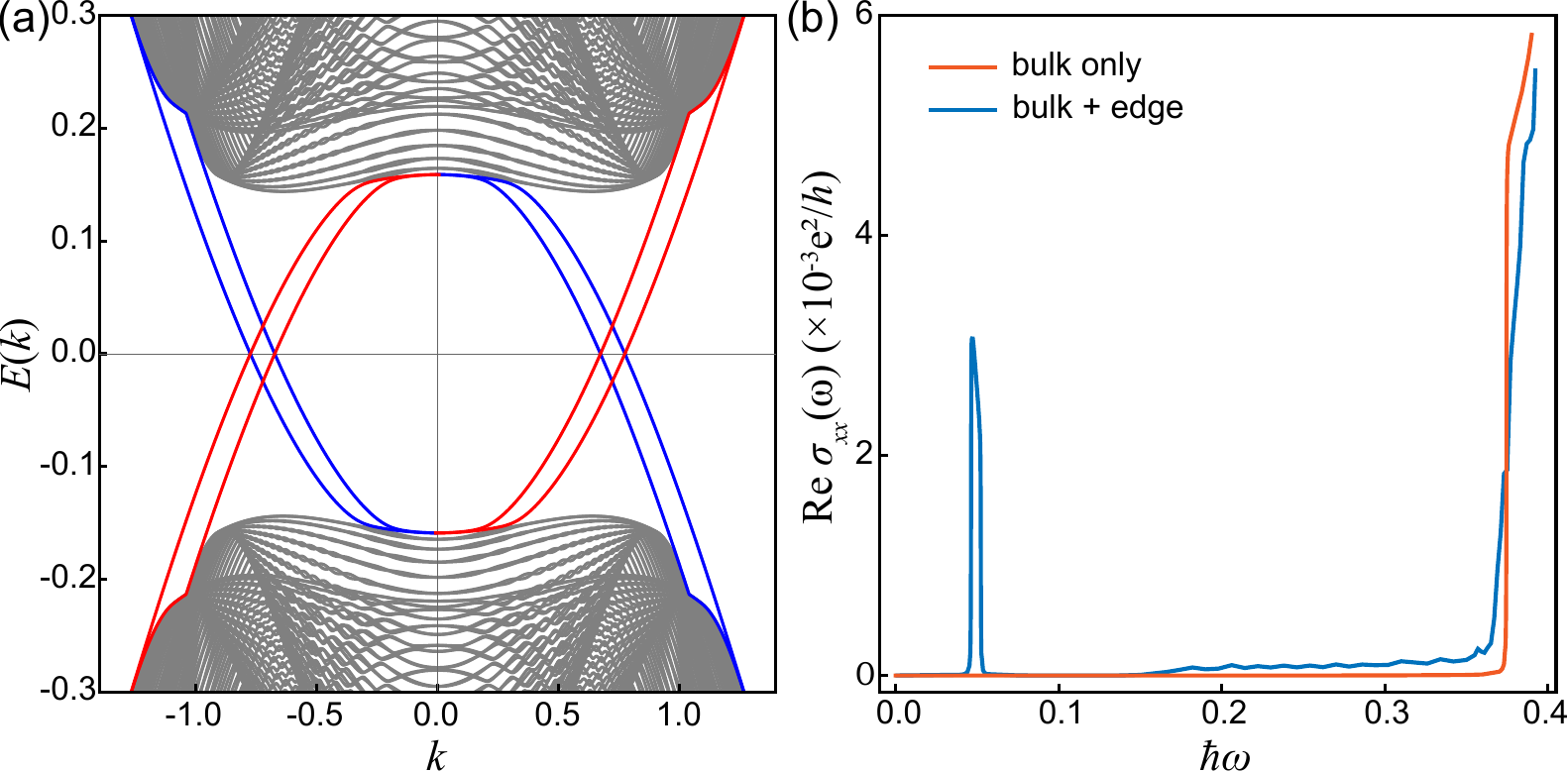}
  \end{center}
  \caption{(a) The energy spectrum of the $d+id$ TSC. (b) The real part of optical conductivity.}
  \label{fig7}
\end{figure}

Since a single band superconductor cannot respond to an electromagnetic field, we consider a double $d+id$-wave TSC. The tight-binding BdG Hamiltonian is:
\begin{equation}
  \begin{aligned}
    H_\text{BdG}(\mathbf{k})= & [4t-\mu-2t \cos(k_x)-2t \cos(k_y)]\tau_3\sigma_0+\lambda[\sin(k_x)\tau_0\sigma_1+\sin(k_y)\tau_3\sigma_2] \\
    & +\Delta[2 (\cos(k_x)-\cos(k_y))\tau_2\sigma_2+2\sin(k_x)\sin(k_y)\tau_1\sigma_2].
  \end{aligned}
\end{equation}
with the basis $\psi_{\mathbf{k}}=(c_{\mathbf{k}\uparrow}, c_{\mathbf{k}\downarrow},c^\dagger_{-\mathbf{k}\uparrow}, c^\dagger_{-\mathbf{k}\downarrow})^T$. $\tau_i$ and $\sigma_i$ are Pauli matrices for Nambu space and spin space, respectively.
The $\lambda$ term represents the spin-orbit coupling, which mixes the spin-up and spin-down bands and thus gives rise to a nonzero coupling parameter $b$ for two Bogoliubov modes \cite{bi2024vertical}.
Here we choose $t = 0.5,~\mu = 0.5,~\lambda = 0.1,~\Delta = 0.2$, and the small positive parameter $\eta$ for calculating conductivity is set as $10^{-4}$.
This model has a BdG Chern number $N=4$, indicating four chiral edge modes on one side, as shown in Fig.~\ref{fig7}(a).
The optical conductivity of this model is presented in Fig.~\ref{fig7}(b).
Similar to the other models in the main text, the lowest transition between the bulk bands is absent.
%The optical transition can only occur between the modes localized at the same edge.
Due to the nature of the gap function, each particle-hole pair is separated in momentum space, thus no intra-pair transition can occur, while inter-pair transitions are allowed. In this model, different pairs of edge modes have the same velocity near zero energy, enabling their optical transition to generate a resonance peak below the bulk gap, as shown in Fig.~\ref{fig7}(b).

\end{document}